\documentclass{article}

\usepackage{arxiv}

\usepackage[utf8]{inputenc} 
\usepackage[T1]{fontenc}    
\usepackage{hyperref}       
\usepackage{url}            
\usepackage{booktabs}       
\usepackage{amsfonts}       
\usepackage{nicefrac}       
\usepackage{microtype}      
\usepackage{lipsum}		
\usepackage{graphicx}
\usepackage{natbib}
\usepackage{doi}
\usepackage{hyperref}
\usepackage{tcolorbox}
\usepackage{listings}
\usepackage{times}
\usepackage{latexsym}
\usepackage{multirow}
\usepackage{amsmath}
\usepackage{makecell}
\usepackage{amssymb}
\usepackage{float}
\usepackage{enumitem}
\usepackage{inconsolata}
\usepackage{arydshln}
\usepackage{multirow}
\usepackage{nicematrix}
\usepackage[capitalise]{cleveref}
\usepackage{pifont}
\usepackage{fontawesome}
\usepackage{textcomp}
\usepackage{soul}
\usepackage{xcolor}

\definecolor{codegreen}{rgb}{0,0.6,0}
\definecolor{codegray}{rgb}{0.5,0.5,0.5}
\definecolor{codepurple}{rgb}{0.58,0,0.82}
\definecolor{backcolour}{rgb}{0.95,0.95,0.92}

\lstdefinestyle{mystyle}{
    backgroundcolor=\color{backcolour},   
    commentstyle=\color{codegreen},
    keywordstyle=\color{magenta},
    numberstyle=\tiny\color{codegray},
    stringstyle=\color{codepurple},
    basicstyle=\ttfamily\footnotesize,
    breakatwhitespace=false,         
    breaklines=true,                 
    captionpos=b,                    
    keepspaces=true,                 
    numbers=left,                    
    numbersep=5pt,                  
    showspaces=false,                
    showstringspaces=false,
    showtabs=false,                  
    tabsize=2
}

\title{Teaching Machines to Code: Smart Contract Translation with LLMs}

\author{ \href{https://orcid.org/0000-0002-6705-6506}{\includegraphics[scale=0.06]{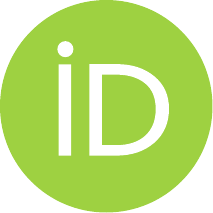}\hspace{1mm}Rabimba Karanjai}\thanks{www.rabimba.me} \\
	Department of Computer Science\\
	University Of Houston\\
	\texttt{rkaranjai@uh.edu} \\
        \And
	Lei Xu \\
 	Department of Computer Science\\
        Kent State University \\
	\texttt{xuleimath@gmail.com}\\
        \And
	Weidong Shi \\
 	Department of Computer Science\\
        University Of Houston \\
	\texttt{wshi3@uh.edu}\\
}

\renewcommand{\shorttitle}{\textit{SolMover}}

\hypersetup{
pdftitle={SMART CONTRACT TRANSLATION WITH LLMS},
pdfsubject={Smart Contract LLM Translation},
pdfauthor={Rabimba Karanjai},
pdfkeywords={Smart Contracts, Machine Learning, Machine Translation, Code Transpilation, LLM, Large Language Model},
}

\begin{document}
\maketitle

\begin{abstract}
	
The advent of large language models (LLMs) has marked a significant milestone in the realm of artificial intelligence, with their capabilities often matching or surpassing human expertise in various domains. Among these achievements, their adeptness in translation tasks stands out, closely mimicking the intricate and preliminary processes undertaken by human translators to ensure the fidelity and quality of the translated content. Despite the advancements in utilizing LLMs for translating programming code across different languages, the domain of smart contract translation, particularly into languages not previously encountered by the LLM, remains largely unexplored. In our research, we present a pioneering approach, SolMover, which harnesses the synergy of two distinct LLMs within a unified framework. This framework is designed to grasp coding principles and apply this understanding to the translation of code into an unfamiliar language. Our study delves into the capacity of LLMs to mimic human learning processes, offering an in-depth evaluation of our methodology for converting smart contracts written in Solidity to Move, a language with limited resources. The framework employs one LLM to decipher coding conventions for the new language, creating a blueprint for the second LLM, which, lacking planning abilities, possesses coding expertise. The empirical evidence from our experiments suggests that SolMover substantially enhances performance compared to gpt-3.5-turbo-1106, and achieves superior results over competitors such as Palm2 and Mixtral-8x7B-Instruct. Additionally, our analysis highlights the efficacy of our bug mitigation strategy in elevating code quality across all models, even outside the SolMover framework.
\end{abstract}


\section{Introduction}
The groundbreaking strides in Large Language Models (LLMs) have been instrumental in shaping a future that mirrors human cognitive abilities, particularly in the understanding and generation of natural language. These models, acclaimed for their near-human intelligence, have significantly advanced the field of natural language processing, as highlighted by several seminal works~\cite{liang2022holistic,bubeck2023sparks,Wu2023ChatGPTOG,moghaddam2023boosting}. The prowess of LLMs in the realm of natural language translation has been especially pronounced, underscoring their capability to execute translations with remarkable accuracy and efficiency~\cite{jiao2023ischatgpt,agrawal-etal-2023-context,zhang2023prompting,vilar2022prompting,lu2023chain}. This progress resonates deeply with the aspirations of early machine translation researchers from the 1960s, prompting a reevaluation of whether LLMs can replicate human translation methodologies~\cite{bar1960demonstration, macklovitch1995future}. Consequently, this beckons the exploration of LLMs in the domain of code-to-code translation, leveraging their advanced translation mechanisms.

In the context of the financial industry, the rapid expansion of decentralized ledger technologies and smart contracts has been noteworthy. The utilization of Uniswap's smart contracts, for instance, achieved an average daily transaction volume of approximately \$7.17 billion in 2021, underscoring the growing significance of smart contracts across various applications, including Confidential Computing and Decentralized Serverless Architectures~\cite{uniswap, 10.1145/3505253.3505259, 10174945}. This surge in adoption raises pertinent questions about the potential of LLMs to not only author smart contracts based on user directives but also ensure their security and robustness.

Our research endeavors to unravel the capabilities and limitations of LLMs in the translation of smart contract code. We investigate the feasibility of employing LLMs to translate smart contracts, focusing on their ability to emulate human translation techniques. This inquiry centers on the potential of LLMs to comprehend a smart contract written in Solidity and to produce an equivalent in another language, Move, in this case. Our approach seeks to encapsulate smart contract concepts within the model, facilitating a universal framework for code transpilation.

To achieve this, we introduce SolMover, a system designed to bridge the gap between Solidity and Move through a multi-step knowledge distillation process. This process involves parsing Solidity files to extract functions and content, subsequently generating high-level tasks that inform the creation of sub-tasks by a fine-tuned LLM. This LLM, having been fine-tuned on textbooks and specifications related to both Solidity and Move, paves the way for a secondary, more specialized code generation model to produce the equivalent Move code. Verification and error correction mechanisms, including move-prover and multi-prompting techniques, further refine the quality of the generated code. Our study aims to elucidate the following research queries:

\textbf{RQ1:} We explore the LLMs' ability to internalize coding concepts, particularly those of the Move language, assessing whether they can learn programmatic rules from textual descriptions and apply these in code generation.

\textbf{RQ2:} We assess the generation of specific sub-tasks from a general prompt, evaluating the utility of these sub-tasks in code production.

\textbf{RQ3:} Our investigation extends to the generation of compilable Move code by a model not extensively trained on Move, questioning the limits of LLMs' adaptability.

\textbf{RQ4:} We examine the efficacy of compiler feedback in debugging and enhancing code quality, and the extent to which prompting techniques aid in this process.

The Move language, developed for the Libra Blockchain, introduces unique asset types with stringent safety measures, posing a unique challenge for code translation from Solidity. This work contributes to the field by demonstrating the potential of LLMs to translate code between languages, refine code generation through compiler feedback, and apply knowledge distillation for translating to low-resource languages. To our knowledge, this is the first study to illustrate the feasibility of using LLMs for generating compilable code in a low-resource language through fine-tuning and strategic model integration.


\definecolor{verylightgray}{rgb}{.97,.97,.97}

\lstdefinelanguage{Solidity}{
	keywords=[1]{anonymous, assembly, assert, balance, break, call, callcode, case, catch, class, constant, continue, constructor, contract, debugger, default, delegatecall, delete, do, else, emit, event, experimental, export, external, false, finally, for, function, gas, if, implements, import, in, indexed, instanceof, interface, internal, is, length, library, log0, log1, log2, log3, log4, memory, modifier, new, payable, pragma, private, protected, public, pure, push, require, return, returns, revert, selfdestruct, send, solidity, storage, struct, suicide, super, switch, then, this, throw, transfer, true, try, typeof, using, value, view, while, with, addmod, ecrecover, keccak256, mulmod, ripemd160, sha256, sha3}, 
	keywordstyle=[1]\color{blue}\bfseries,
	keywords=[2]{address, bool, byte, bytes, bytes1, bytes2, bytes3, bytes4, bytes5, bytes6, bytes7, bytes8, bytes9, bytes10, bytes11, bytes12, bytes13, bytes14, bytes15, bytes16, bytes17, bytes18, bytes19, bytes20, bytes21, bytes22, bytes23, bytes24, bytes25, bytes26, bytes27, bytes28, bytes29, bytes30, bytes31, bytes32, enum, int, int8, int16, int24, int32, int40, int48, int56, int64, int72, int80, int88, int96, int104, int112, int120, int128, int136, int144, int152, int160, int168, int176, int184, int192, int200, int208, int216, int224, int232, int240, int248, int256, mapping, string, uint, uint8, uint16, uint24, uint32, uint40, uint48, uint56, uint64, uint72, uint80, uint88, uint96, uint104, uint112, uint120, uint128, uint136, uint144, uint152, uint160, uint168, uint176, uint184, uint192, uint200, uint208, uint216, uint224, uint232, uint240, uint248, uint256, var, void, ether, finney, szabo, wei, days, hours, minutes, seconds, weeks, years},	
	keywordstyle=[2]\color{teal}\bfseries,
	keywords=[3]{block, blockhash, coinbase, difficulty, gaslimit, number, timestamp, msg, data, gas, sender, sig, value, now, tx, gasprice, origin},	
	keywordstyle=[3]\color{violet}\bfseries,
	identifierstyle=\color{black},
	sensitive=true,
	comment=[l]{//},
	morecomment=[s]{/*}{*/},
	commentstyle=\color{gray}\ttfamily,
	stringstyle=\color{red}\ttfamily,
	morestring=[b]',
	morestring=[b]"
}

\lstset{
	language=Solidity,
	backgroundcolor=\color{verylightgray},
	extendedchars=true,
	basicstyle=\footnotesize\ttfamily,
	showstringspaces=false,
	showspaces=false,
	numbers=left,
	numberstyle=\footnotesize,
	numbersep=9pt,
	tabsize=2,
	breaklines=true,
	showtabs=false,
	captionpos=b
}

\section{SolMover}
In the present section, we unveil the architecture of the SolMover framework.

\begin{figure}
\centering
\includegraphics[width=1\linewidth,height=5.8in]{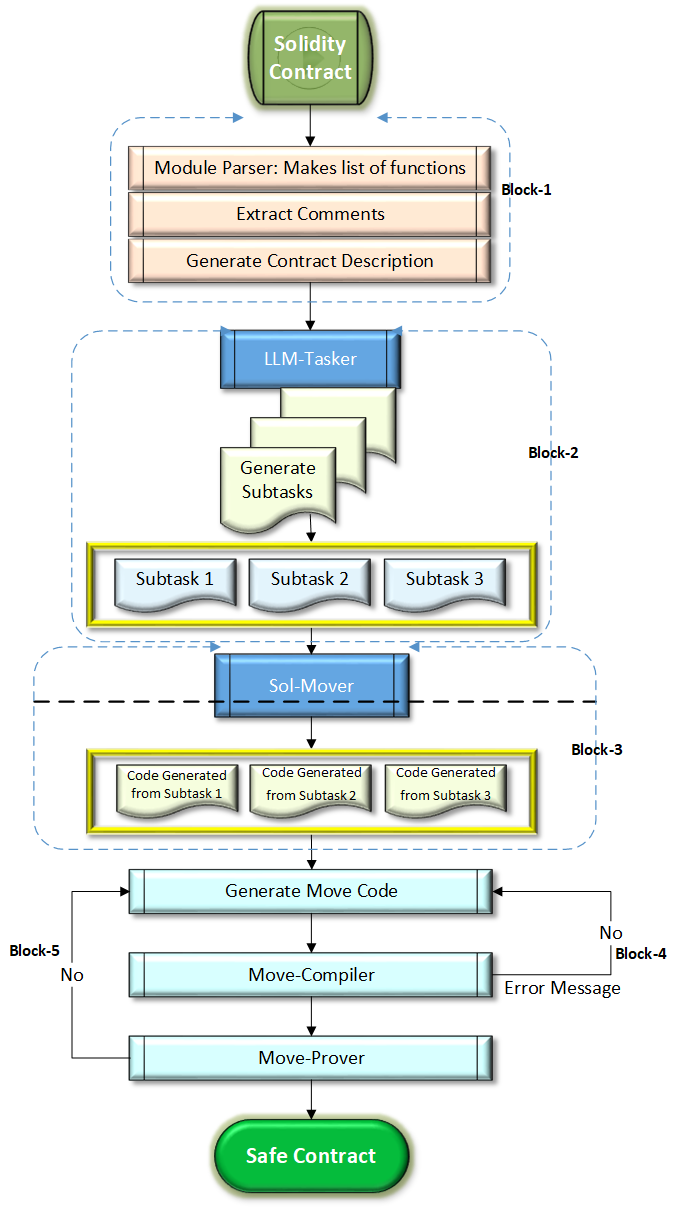}
\caption{The architecture of the \textbf{SolMover} framework is delineated across five phases: (1) Task Creation involves the parsing of a Solidity file to generate the initial task prompt, utilizing comments and keywords found within the smart contract. (2) In Concept Mining, a retrieval-augmented technique is employed to extract concepts from Move programming resources to create sub-tasks derived from the initial prompt. (3) Code Generation leverages these sub-tasks, refined through concept mining, to produce code for each sub-task. These segments are subsequently compiled into a unified Move file, representing the candidate translation. (4) This code undergoes compilation; failures in this process trigger a feedback loop, where the error is reported back to SolMover for prompt modification and code regeneration. (5) Successfully compiled code is subjected to verification through the Move Prover to assess its provability. Should verification fail, the error is integrated into the prompt for corrective regeneration by SolMover.}
\label{fig:solmover}
\end{figure}

As illustrated in \figurename~\ref{fig:solmover}, the SolMover framework is composed of several interconnected modules. \textit{In the first module}, the analysis of the Solidity smart contract slated for translation is carried out, employing a method akin to that proposed by Karanjai et al. \cite{10316829} for the extraction of functions and comments from the Solidity file. The module synergizes a parser and a comment extractor to mine comments from the code, utilizing a prompt template similar to the one described in \cite{10316829} to formulate the initial task prompt. \textit{The second module} features the LLM tasker, which sources from Move documentation, whitepapers, and tutorials to conduct a Retrieval Augmented Search, generating sub-tasks aligned with the initial prompt. These sub-tasks are subsequently processed by a second, specifically tuned LLM (Sol-Mover), culminating in the assembly of the final Move code in \textit{Module 4}. Here, the move-cli tool attempts to compile the translated code, recording successful compilations and flagging compilation failures by feeding error messages back to Sol-Mover for iterative code regeneration, with a maximum of five attempts before declaring a translation failure. \textit{In the fifth module}, the framework endeavors to validate the compilable code's correctness through the Move Prover \cite{10.1007/978-3-030-53288-8_7}, aiming to formally prove the smart contract's adherence to its specifications. Failures in verification prompt a feedback loop, where Sol-Mover reattempts correction by incorporating the identified errors into the prompt. After a maximum of five iterations, if the code's correctness is established, it is designated as a safe contract; otherwise, it is considered compilable but unverified.

\subsection{Code Understanding}
As we transition our efforts from code translation to code generation, it becomes imperative to comprehend the functionalities and intentions encapsulated within the original input code. To facilitate this understanding, we have devised a parser dedicated to analyzing Solidity files. This parser is capable of extracting comments, functions, and formulating an initial candidate prompt based on the content of the Solidity file, despite the description being \textcolor{red}{incomplete}.

The significance of \textit{comments} in deciphering the functionality and intended outcomes of a Solidity smart contract cannot be overstated. These comments, embedded within the code, serve as guides, illuminating the underlying logic and primary objectives of the contract. They are instrumental in grasping the developer's original intentions and guaranteeing the precise execution of the code. The parsing of Solidity code to extract these insights is executed utilizing methodologies paralleled to those described in~\cite{karanjai2023smarter}.

The \textit{Topic} of the Solidity code essentially encapsulates the contract's functionality, typically inferred from the preliminary comments and the functions utilized within the code. This is exemplified in the code snippet presented in Listing \ref{lst:solidty}, which outlines a rudimentary hotel and vending machine Solidity smart contract. This contract enables the renting of hotel rooms, allowing payments for a room under the condition it is vacant. Upon payment, funds are transferred to the owner's account. This contract metaphorically resembles a vending machine, where an input of funds, upon successful validation, results in the delivery of a product or service, akin to a gumball machine. The concept includes potential expansions such as unlocking doors or dispensing key codes post-payment, demonstrating the contract's foundational logic and its intended purpose.

\subsection{Concept Mining}\label{concept}
Drawing inspiration from the principles underlying human translation methodologies as explicated by Gile~\cite{gile2009basic}, the procedure of concept mining necessitates that the Large Language Model (LLM) initially generates output that is congruent with the pertinent concepts and knowledge essential for the task of transpilation.

The process of concept mining entails the fine-tuning of an LLM using textbook data specific to the Move programming language. Following the methodology outlined by Gunasekar et al~\cite{gunasekar2023textbooks}, the dataset is meticulously prepared and the model is subsequently fine-tuned. Furthermore, strategies for identifying relevant textual material are incorporated as per the guidelines provided by Zhang et al~\cite{zhang2023reaugkd} and Lu et al~\cite{lu2022reacc}. The dataset preparation is depicted in \figurename~\ref{fig:conceptdist}.

In this context, various sources are employed for the extraction of concepts, including textbooks on Move \cite{IntroductionTheMoveBook-2023-08-12,TheMoveLanguageTheMoveBook-2022-11-25}, blogs pertaining to Move programming \cite{movepatterns}, and tutorials featuring code snippets \cite{movelanguagedocumentationtutorialatmainmovelanguagemoveGitHub-2023-12-19}, alongside code samples from the repository~\cite{SuiBasicsSuiMovebyExample-2023-03-27}. Notably, the fine-tuning of the model intentionally eschews the use of any code corpus.

For preparing the dataset we use scripts from \cite{hfcodegendatapreparedatasetpyatmainsayakpaulhfcodegenGitHub-2023-12-19} modified to work for our files. 

\subsubsection{Architecture}\label{Architecture}
Within our architectural framework, we incorporate the methodology of Retrieval Augmented Generation as delineated by Zhang et al~\cite{zhang2023reaugkd}, tailored to our specific needs for concept identification. This adaptation acknowledges the distinct requirements of our task compared to the original application of the method.

We posit the existence of a comprehensive textbook database, replete with a diverse array of textual content and code snippets, as detailed in Section \ref{concept}. From this repository, we derive $D$ text snippets by segmenting the documents. Given that these documents are in HTML format, segmentation is achieved through the identification of \textbf{H} tags or \textbf{Headings} in the HTML code, or alternatively, in the absence of prominent text blocks, accompanied by smaller textual segments. This strategy aims to delineate concept headings and their corresponding descriptions. Utilizing the Dense Passage Retriever (DPR) methodology~\cite{karpukhin2020dense}, we further divide these text segments into equally sized fragments, serving as the primary units for search and retrieval. This approach is predicated on the premise that such a division facilitates enhanced retrieval efficacy, as suggested by Karpukhin et al~\cite{karpukhin2020dense}, and concurrently, augments the scalability of our framework to accommodate substantial file sizes. Consequently, we amass $M$ text fragments constituting the retrieval database $C = {c_1, c_2, ..., c_M}$.

Given a set of query keywords $X = {x_1, x_2, ..., x_k}$, a search mechanism $\textbf{R}: (X,C) \rightarrow C$ is employed to locate the most akin text fragment $c_s$ within $C$. Subsequently, a generator $\textbf{G}$ is tasked with predicting the ensuing text tokens $Y = {x_{k+1}, ..., x_{k+n}}$, where $n=1$ denotes the granularity of token-level text prediction, guided by both the context and the retrieved text.

This process is formally encapsulated as $P(Y)=\prod_{i=1}^{n} P(x_{k+i}|c_s,x_{1:k+i-1})$.

For the search functionality, we employ the BM25 algorithm~\cite{robertson2009probabilistic}, akin to its utilization within ElasticSearch frameworks. BM25 operates as a term-based search methodology, employing a bag-of-words model to ascertain lexical similarity between the query and document fragments. The DPR framework~\cite{karpukhin2020dense} is constituted by two bidirectional transformers, $E_C$ and $E_Q$, where $E_C$ is responsible for encoding and indexing each segmented text within the database $C$. The encoded representation of the \texttt{[CLS]} token is extracted as the output, facilitating the computation of similarity as $sim(q, c) = E_C(c)^TE_Q(q)$. In the training phase, we adopt the batch negative strategy for loss computation, as illustrated by van den Oord et al~\cite{oord2018representation}. Contrary to the approach of Lu et al~\cite{lu2022reacc} which overlooks hard negatives, our methodology aligns with that of Karpukhin et al~\cite{karpukhin2020dense}, embracing a comprehensive perspective on negative sampling.

\begin{equation}
\begin{aligned}
&L(q, c^+, c^-_1, c^-_2, ..., c^-m) \ =
&-log\frac{e^{sim(q, c^+)}}{e^{sim(q, c^+)} + \sum^m{i=1}e^{sim(q, c^-_i)}}
\end{aligned}
\end{equation}

\begin{figure}
\centering
\includegraphics[width=1\linewidth]{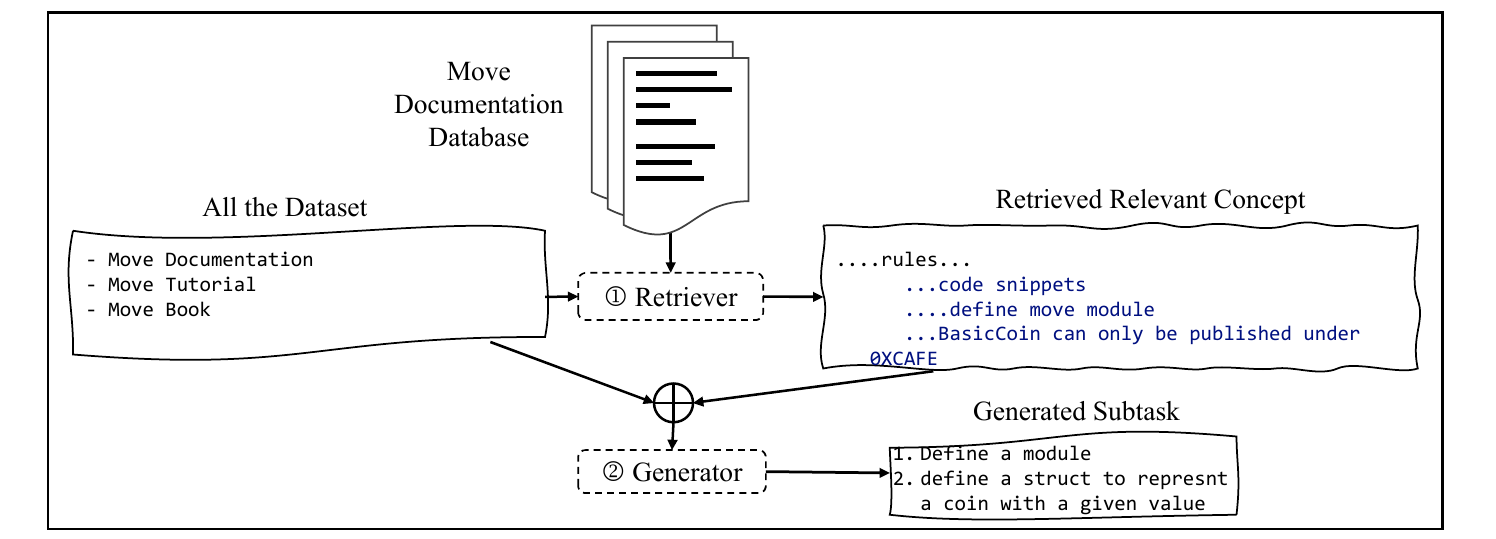}
\caption{The process of concept selection.}
\label{fig:concept1}
\end{figure}

\subsection{Concept Selection}
The process of Concept Selection signifies the culmination of our sub-task generation phase. While keywords, topics, and demonstrations of relevant text are instrumental in generating sub-tasks for our translation endeavor, it is crucial to acknowledge that not all generated outcomes may serve a practical purpose. Specifically, the LLM might engender sub-tasks predicated on either trivial or extraneous content, potentially leading the Sol-Mover LLM to produce either superfluous code snippets or, in less favorable instances, incorrect code. To mitigate such outcomes, we refine our sub-task generation to encompass only those concepts accompanied by suitable code snippets. These code snippets, in turn, are incorporated into the Sol-Mover as integral components of the sub-task prompts, thereby enhancing the LLM's capability to generate pertinent sub-tasks, as illustrated in Figure \ref{fig:concept1}.

\subsection{Sol-Mover Code Generator}\label{SolGen}

Sol-Mover emerges as our dedicated Move code generator, which processes the sub-tasks delineated by the preceding LLM, as depicted in Figure \ref{fig:concept1}, to produce Move code. The foundation of Sol-Mover is an instruction-trained model, derived from Alpaca \cite{alpaca}, an open-source LLM refined through fine-tuning on the LLaMA \cite{touvron2023llama} architecture. This model incorporates 52k instances of Self-Instruct data alongside user-contributed dialogues from ShareGPT \cite{sharegpt2023}. Utilizing langchain, a prompting infrastructure is devised, assimilating each sub-task in conjunction with any associated code snippets to facilitate code generation via the fine-tuned model. Subsequent to the generation phase, the responses are amalgamated into a unified file, adhering to the task sequence, thereby reconstructing the translated code pertaining to the original task. This consolidated code is then subjected to compilation via move-cli. Compilation errors, if any, are meticulously logged and integrated as experiential data within the langchain agent, which is subsequently relayed back to Sol-Mover for corrective measures. This iterative process is executed up to five times. Upon successful passage through the compilation phase, the code translation is deemed successful. However, the verification of its correctness remains, prompting the involvement of Move Prover, an automated formal verification tool for Move code. Any discrepancies identified by the Prover are directed back to Sol-Mover, initiating up to five iterations of corrective adjustments. The intricacies surrounding the training of both Sol-Mover and the sub-task creation LLM are expounded in Section \ref{techspec}.

\subsection{Dataset for Solidity as Input Candidate}\label{soldataset}
In our research, we encountered a significant challenge due to the inability to employ pre-existing datasets like HumanEval~\cite{chen2021codex}, which predominantly cater to Python programming. This limitation necessitated the generation of a new dataset tailored to our specific requirements. To ensure the accuracy of code validation, our methodology involved a rigorous selection process for source files, prioritizing those available on GitHub under open-source licenses that offer permissive usage terms.

To systematically compile Solidity code from various projects, we developed a scraper utilizing GitHub's APIs. The criteria for corpus collection were meticulously defined to enhance the quality and relevance of the dataset:

\begin{itemize}
\item We targeted projects that had garnered a minimum of 50 GitHub stars, a benchmark indicative of a project's interest, popularity, and its appeal to a human audience.
\item The explicit tagging of a project with "Solidity" as a programming language was a prerequisite, ensuring the relevance of the collected code.
\item A substantial volume of comments within the projects was deemed crucial. This was to facilitate the generation of scalable prompts and to lay the groundwork for an effective experimental pipeline.
\end{itemize}

Subsequent to this targeted collection, an additional layer of processing was applied to sift through the gathered materials, excluding any files not authored in Solidity.

\paragraph{Compilation of the raw code corpus.}
Our approach leveraged GitHub, a rich repository of publicly accessible source code, to mirror repositories prominently featuring the Solidity language tag and possessing at least 50 stars. This procedure enabled the extraction of Solidity files, which were then aggregated to form the base dataset intended for training purposes.

\section{Technical Specification}\label{techspec}

In our research, we delineate the training regimen and model selection strategies for our models, focused on Concept understanding and sub-task generation, as well as the fine-tuning of Alpaca for code generation tasks.

\subsection{Retrieval Augmented LLM for Concept Understanding}

In the pursuit of facilitating code-to-code translation for languages with very limited resources such as Move, our framework addresses a crucial challenge: premise selection~\cite{urban2004mptp}. Whereas existing frameworks based on Large Language Models (LLMs) for knowledge distillation~\cite{zhang2023reaugkd} and code completion~\cite{robertson2009probabilistic} predominantly generate subsequent steps predicated on the immediate preceding input, the creation of coherent plans and sub-tasks is significantly reliant on premises. These include rules and code examples extracted from related descriptions, which are essential for understanding and translating code accurately.

The integration of all possible contexts within the limited input window of an LLM is infeasible, and attempts to bypass this limitation via LangChain's agent-based memory have not yielded substantially improved outcomes. Traditional methods are constrained to memorize associations between the current task state and the subsequent relevant task, contingent on the original task. This approach is effective when the training data encompasses similar problems; however, for low-resource languages like Move, this methodology encounters limitations. Moreover, since Move is syntactically akin to Rust, pre-trained models on Rust may inadvertently introduce incorrect premise candidates. This issue is exacerbated in scenarios requiring the generation of code for tasks that were not encountered in the training dataset.

To address these challenges, our framework enhances LLM memorization with a strategy for explicit context selection, thereby facilitating improved concept identification. This is achieved by extracting contexts directly from texts where they are defined and utilized, thereby enabling the identification of pertinent contexts through the augmentation of LLMs with retrieval capabilities.

Furthermore, it is imperative to constrain context retrieval to a manageable number of contexts to ensure its efficacy. As outlined in Section \ref{Architecture}, our retrieval mechanism is based on the Dense Passage Retriever (DPR) but incorporates two significant modifications tailored to our specific requirements. Firstly, the scope of context available for retrieval is deliberately restricted by initially extracting comments and functions, which serve as keywords in the initial prompt. This reduction simplifies the retrieval task by approximately 75\%. Secondly, unlike the conventional DPR that benefits from the inclusion of hard negatives (irrelevant contexts that are challenging to differentiate) in its training, our approach employs in-file hard negatives. This method utilizes negative contexts derived from within the prompt itself, enhancing the model's ability to discern relevant from irrelevant contexts effectively.

\subsubsection{Experimental Setup}
The training is in two stages. In the first stage, we train the retriever and use it to retrieve 100 context states for all contexts for each task. Second, we train the sub-task generator, taking as input the retrieved context. We evaluate the task generator by combining it with the best-first search to generate valid sub-tasks. Training takes 7 days on a single NVIDIA A100 GPU. 

\subsection{Sol-Mover Code Generation Training}

For the development of our code generator, we employed the Alpaca model and subjected it to a fine-tuning process utilizing code examples derived from the instruction training dataset, as elaborated upon in Section \ref{SolGen}. The corpus for fine-tuning was notably small, comprising a select compilation of code samples. This curated selection included examples of Fungible Tokens~\cite{funtoken,basiccoin,diem, token,gas}, Non-Fungible Tokens (NFTs)\cite{nft,merklenft}, and Decentralized Finance (DeFi) applications\cite{defi,coinswap,starswap}, which served as a foundation for the dataset's creation.

The primary objective of this fine-tuning process was to imbue the Alpaca model with the capability to form associations with Move code constructs. By integrating these specific code examples into the training regimen, the model was tailored to recognize and generate Move code, facilitating the creation of more accurate and contextually relevant code outputs. This approach ensured that Alpaca could leverage its understanding of the underlying concepts and patterns within Move programming, thereby enhancing its efficiency and effectiveness as a code generator within our framework.

\section{Experiments}\label{experiment}

\subsection{Experimental Setup}

\paragraph{Models.}
In our exploration, we evaluated four Large Language Models (LLMs), including both open and closed-source models at the forefront of current technology.

gpt-3.5-turbo-1106: This model represents OpenAI's implementation of the Reinforcement Learning from Human Feedback (RLHF) technique~\cite{ouyang2022training}, showcasing remarkable capabilities. Access to this closed-source language model is facilitated through OpenAI's official API.
Alpaca: Developed from the LLaMA model~\cite{touvron2023llama1}, Alpaca is an open-source model refined to follow instructions. It has been enhanced with a dataset comprising 52,000 Self-Instruct examples~\cite{selfinstruct}, showcasing its adaptability to diverse tasks.
Mixtral: The Mixtral-8x7B-Instruct variant, based on the Mixtral-8x7B architecture~\cite{mixtral_moe}, stands out for its instruction-following prowess and creative output capabilities. Available for public use, this model is adept at generating text across different formats, language translation, answering questions, and creating varied forms of creative content.
Palm2: Google's PaLM 2 model is distinguished by its advanced language understanding and reasoning capabilities. Excelling in complex tasks involving code, mathematics, and multilingual translation, PaLM 2 leverages its substantial size and innovative techniques to achieve superior performance.
For our experimentation with Alpaca, specifically the 7B model variant, inference was executed on a single NVIDIA V100 32GB GPU.

\paragraph{Comparative Methods}
In the context of our code generation endeavor, we confined our methodology to the single-candidate approach. This decision was informed by our framework's reliance on a sequential response mechanism, necessitating the focus on only the initial candidate response.

Within single candidate method, we consider:
\begin{itemize}[leftmargin=10pt]
    \item \textbf{Baseline}: Zero-shot smart contract translation with temperature set to 0 (default for remainder of experiments in this paper).
    \item \textbf{5-Shot} ~\cite{hendy2023good}: Prompting the model with five exemplars of superior quality, specifically selected from the sub-task domain and prepended to the test input, has been demonstrated to yield optimal overall performance, as per ~\cite{hendy2023good}. Further augmenting the number of examples, however, does not appear to produce any appreciable enhancement in outcomes.
\end{itemize}

\paragraph{Metrics and Benchmark}
In our exploration of code translation challenges for the notably resource-scarce language Move, we encountered the absence of pre-existing benchmarks suitable for code generation or testing evaluations. This limitation is compounded by the distinctive nature of Move smart contracts, which complicates the creation of a segregated evaluation dataset for training and testing purposes. Additionally, the specific requirements of a Code Translation Task, as opposed to straightforward code generation, render it impractical to rely solely on public code repositories like GitHub for the procurement of precise code examples for comparative analysis.

Given these constraints, we have adopted a set of evaluation criteria tailored to the unique context of our study:
\begin{itemize}
    \item Code Compilability of the translated smart contract
    \item Code Correctness of the translated smart contract
    \item Bug mitigation of first candidate translation
\end{itemize}

Based on this we evaluate the effectiveness of our approach.

\section{Results}
We have run the experiments on all four LLMs defined in Section \ref{experiment}. However, both Palm2 and Mixtral did not produce any tangible output for any of our candidate translations as we can see from Table \ref{compilable}.

\begin{table}[]
\resizebox{\columnwidth}{!}{%
\begin{tabular}{@{}|l|l|l|@{}}
\toprule
LLMs                                      & Compilable Code? & Performance                                          \\ \midrule
gpt-3.5-turbo-1106 & Y                & Mixed                                                \\ \midrule
Solmover (Two   LLM Combined)             & Y                & Mixed                                                \\ \midrule
Mixtral-8x7B-Instruct                     & N                & Mixes different languages and generate unusable code \\ \midrule
LLama2                                    & N                & Mixes different languages and generate unusable code \\ \midrule
Palm2                                     & N                & For Move only generates code snippets and plan       \\ \bottomrule
\end{tabular}%
}
\caption{\label{compilable}Code Translation Capability of Four LLMs.}
\end{table}

Henceforth we will only report the results found in our SolMover framework and gpt-3.5-turbo-1106. 

\subsection{Successful Smart Contract Translation}

We ran a total of 734 Solidity contracts from our collected dataset in Section \ref{soldataset} through both gpt-3.5-turbo-1106 and SolMover. Each model was allocated an identical quantity of tasks totaling 734. The performance is evaluated based on successful compilation (SC), improvements post-error feedback, and further enhancements after Move Prover feedback. The results are detailed in Table \ref{translate1}.

\begin{table}[]
\resizebox{\columnwidth}{!}{%
\begin{tabular}{@{}|l|l|l|l|l|@{}}
\toprule
LLM                                       & Total Translation Task & Successful Compilation(SC) & SC After Error Feedback & SC after Move Prover Feedback \\ \midrule
{\color[HTML]{353740} gpt-3.5-turbo-1106} & 734                    & 204                        & 229                     & 229                           \\ \midrule
Solmover (Two   LLM Combined)             & 734                    & 313                        & 397                     & 401                           \\ \bottomrule
\end{tabular}%
}
\caption{\label{translate1}Successful Translations.}
\end{table}

\begin{figure}
    \centering
    \includegraphics[width=1\linewidth]{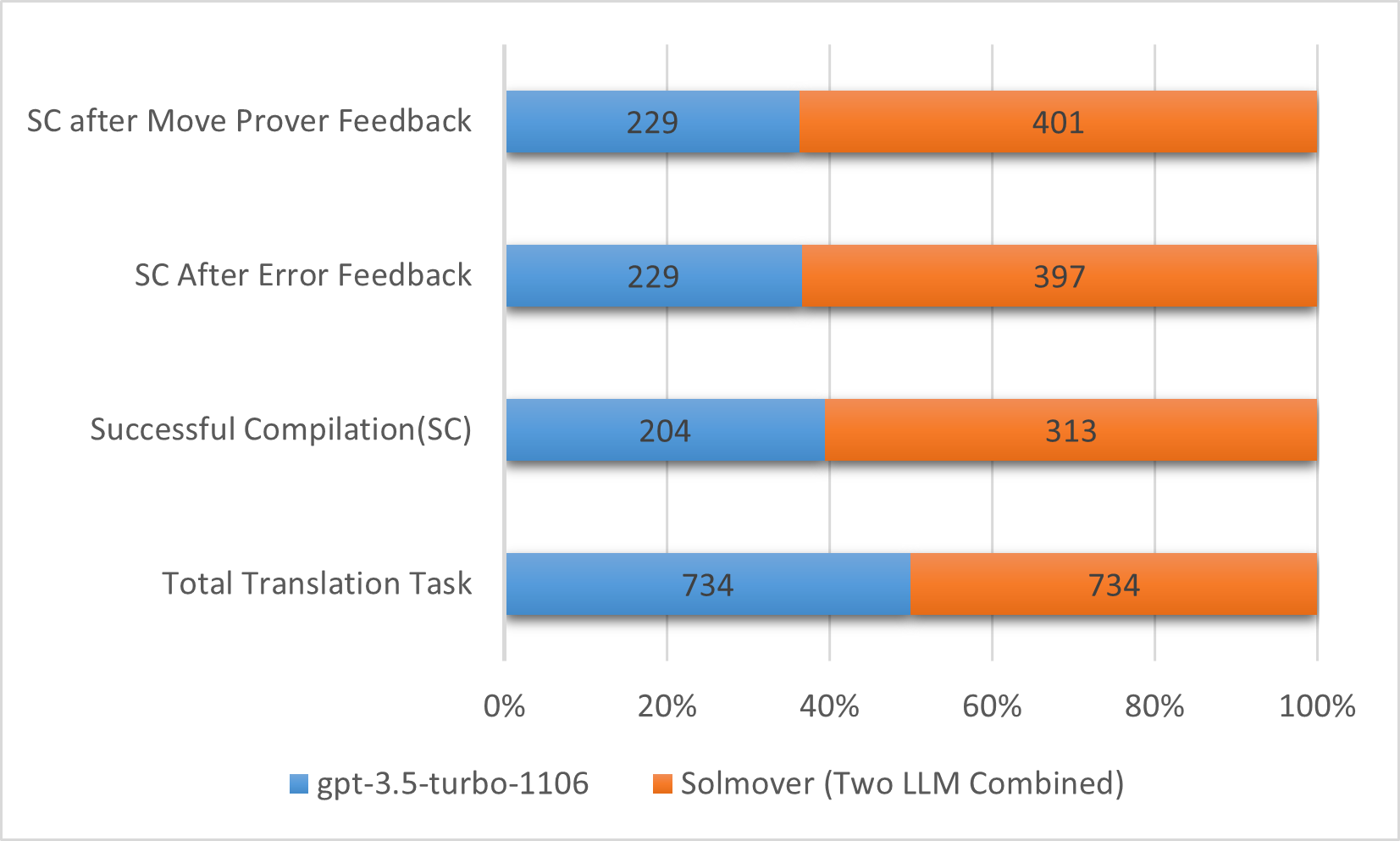}
    \caption{Successful Translations.}
    \label{fig:fig:tran1}
\end{figure}

As we can see from  \figurename~\ref{fig:fig:tran1}. 

\begin{itemize}
    \item Initial SC rates were 204 for \texttt{gpt-3.5-turbo-1106} and 313 for \texttt{Solmover}.
    \item Post error feedback, SC rates improved to 229 for \texttt{gpt-3.5-turbo-1106} and significantly to 397 for \texttt{Solmover}.
    \item Subsequent to Move Prover feedback, \texttt{gpt-3.5-turbo-1106} remained static at 229 SCs, whereas \texttt{Solmover} exhibited a marginal increase to 401 SCs.
\end{itemize}

In this comparative analysis, the combined language model \texttt{Solmover} exhibited superior performance over \texttt{gpt-3.5-turbo-1106} in a translation task involving 734 items. \texttt{Solmover} outperformed in successful compilations initially, after error feedback, and after Move Prover feedback, with the latter showing no improvement at the final stage. This indicates that the integration of two language models in \texttt{Solmover} significantly enhances its ability to learn from feedback and improve its output, making it more adept at handling complex code translation tasks.

\subsection{Reducing Bugs using Iterative Error Feedback}

As we described in Section \ref{SolGen} we try to mitigate bugs introduced by the LLMs for code generation by iterative prompting with error code as a guide. This compiler error-based prompting iterative loop runs at a maximum of five times before marking the translation case as unsuccessful. We run this experiment on both of the LLMs to decouple the concept mining part from the bug mitigation part, and to see if just iterative prompting alone can help achieve better performance on state-of-the-art commercial LLMs too. We report the result in the Table \ref{translate2}. The focus here is on incomplete translations (IC) before and after error feedback is applied.

\begin{table}[]
\resizebox{\linewidth}{!}{%
\begin{tabular}{@{}|l|l|l|l|@{}}
\toprule
LLM                                       & Total Translation Task & Incomplete Translation (IC) & IC After Error Feedback \\ \midrule
{\color[HTML]{353740} gpt-3.5-turbo-1106} & 734                    & 187                         & 110                     \\ \midrule
Solmover (Two   LLM Combined)             & 734                    & 201                         & 134                     \\ \bottomrule
\end{tabular}%
}
\caption{\label{translate2}Bug Mitigation.}
\end{table}

Here we first log the incomplete translations without any kind of compiler feedback. Then we log how both compiler feedback and move provers feedback affect the LLM's bug mitigation capabilities.

\begin{figure}
    \centering
    \includegraphics[width=1\linewidth]{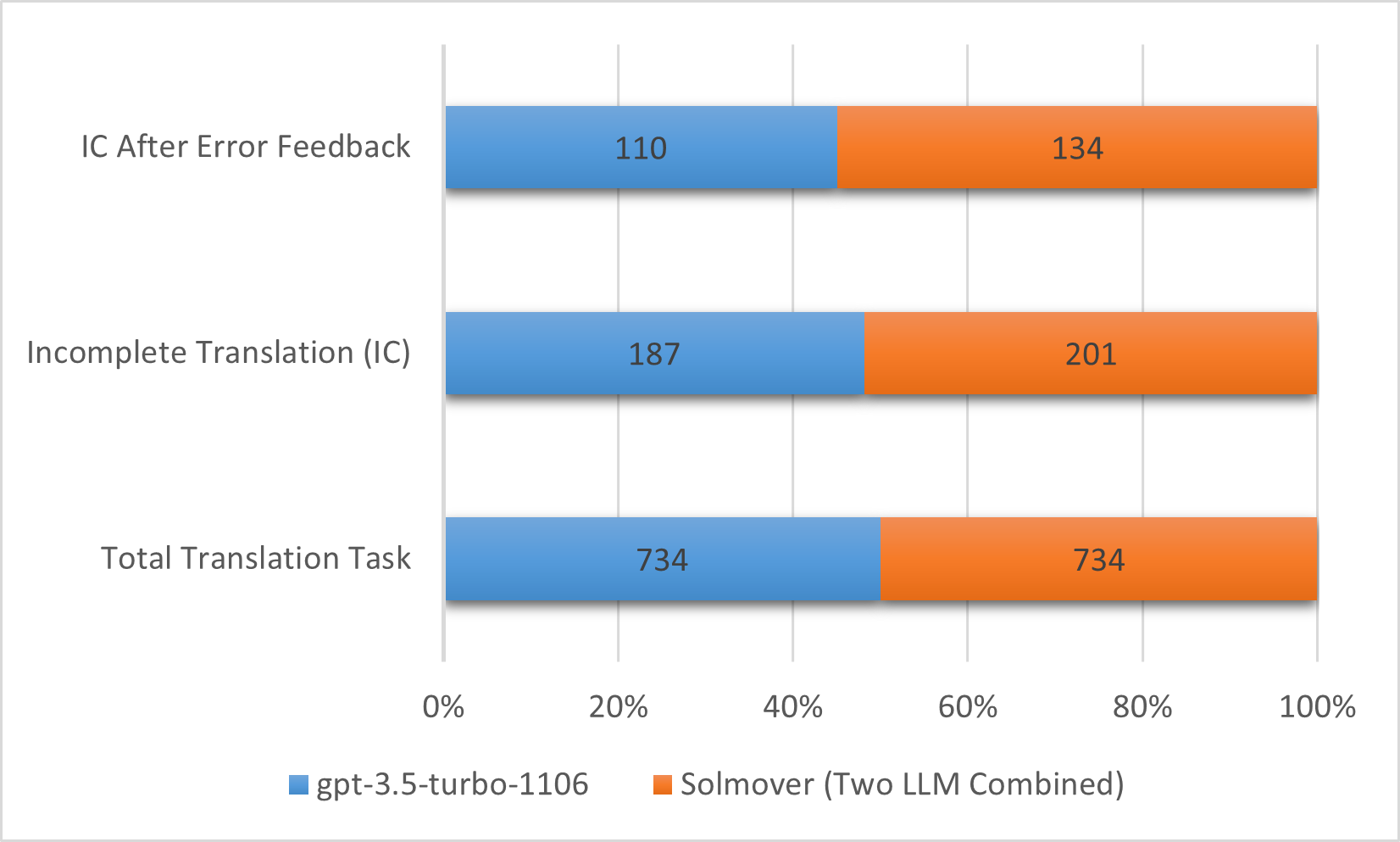}
    \caption{Bug Mitigation.}
    \label{fig:tran2}
\end{figure}

As we can see from Figure \ref{fig:tran2} the results indicate that both models benefited from error feedback, with \texttt{gpt-3.5-turbo-1106} showing a more substantial reduction in Incomplete Completions. Producing more compilable code after the feedback loop. Despite \texttt{Solmover} starting with more Incomplete Completions(ICs), it did not reduce its ICs as effectively as \texttt{gpt-3.5-turbo-1106} post-feedback.

In the comparison of incomplete translations for two language models, \texttt{gpt-3.5-turbo-1106} and \texttt{Solmover (Two LLM Combined)}, both began with a similar number of tasks. \texttt{gpt-3.5-turbo-1106} initially reported fewer incomplete translations and also showed greater improvement after receiving error feedback. This suggests that while \texttt{Solmover} had a higher starting point of incomplete tasks, it was less responsive to feedback in reducing these errors compared to \texttt{gpt-3.5-turbo-1106}.

\begin{itemize}
    \item Pre-feedback, \texttt{gpt-3.5-turbo-1106} had 187 ICs, while \texttt{Solmover} had slightly more with 201 ICs.
    \item Post-error feedback, ICs reduced to 110 for \texttt{gpt-3.5-turbo-1106} and to 134 for \texttt{Solmover}.
\end{itemize}
\subsection{Correctness}
The evaluation of code correctness in our study leverages the capabilities of the Move Prover, a theorem prover specifically designed for the Move programming language. The criterion for determining the safety of a contract hinges on whether the Move Prover can successfully verify the contract without encountering any errors. Contracts that are verifiable by the Move Prover are classified as safe. Conversely, if the Move Prover identifies errors, rendering it unable to prove the contract's correctness, such contracts are categorized under the status of correctness not proved.

To further refine our understanding of code correctness and its implications across different compilers, we employ a similar iterative methodology as used in the compilation error feedback loop. This approach enables us to assess the impact of iterative feedback on the performance of various compilers in terms of enhancing the correctness of the translated code.

The outcomes of this comprehensive evaluation process, focusing on the correctness of translated Move contracts and their verification status as assessed by the Move Prover, are meticulously compiled and presented in Table \ref{translate3}. This systematic analysis aims to shed light on the effectiveness of our code translation framework in producing functionally accurate and provably safe smart contracts within the Move ecosystem.

\begin{table}[]
\resizebox{\columnwidth}{!}{%
\begin{tabular}{@{}|l|l|l|l|@{}}
\toprule
LLM                                       & Successful Compilation(SC) & SC After Error Feedback & SC after Move Prover Feedback \\ \midrule
{\color[HTML]{353740} gpt-3.5-turbo-1106} & 204                        & 229                     & 229                           \\ \midrule
Solmover (Two   LLM Combined)             & 313                        & 397                     & 401                           \\ \bottomrule
\end{tabular}%
}
\caption{\label{translate3}Contract Correctness.}
\end{table}

\begin{figure}
    \centering
    \includegraphics[width=1\linewidth]{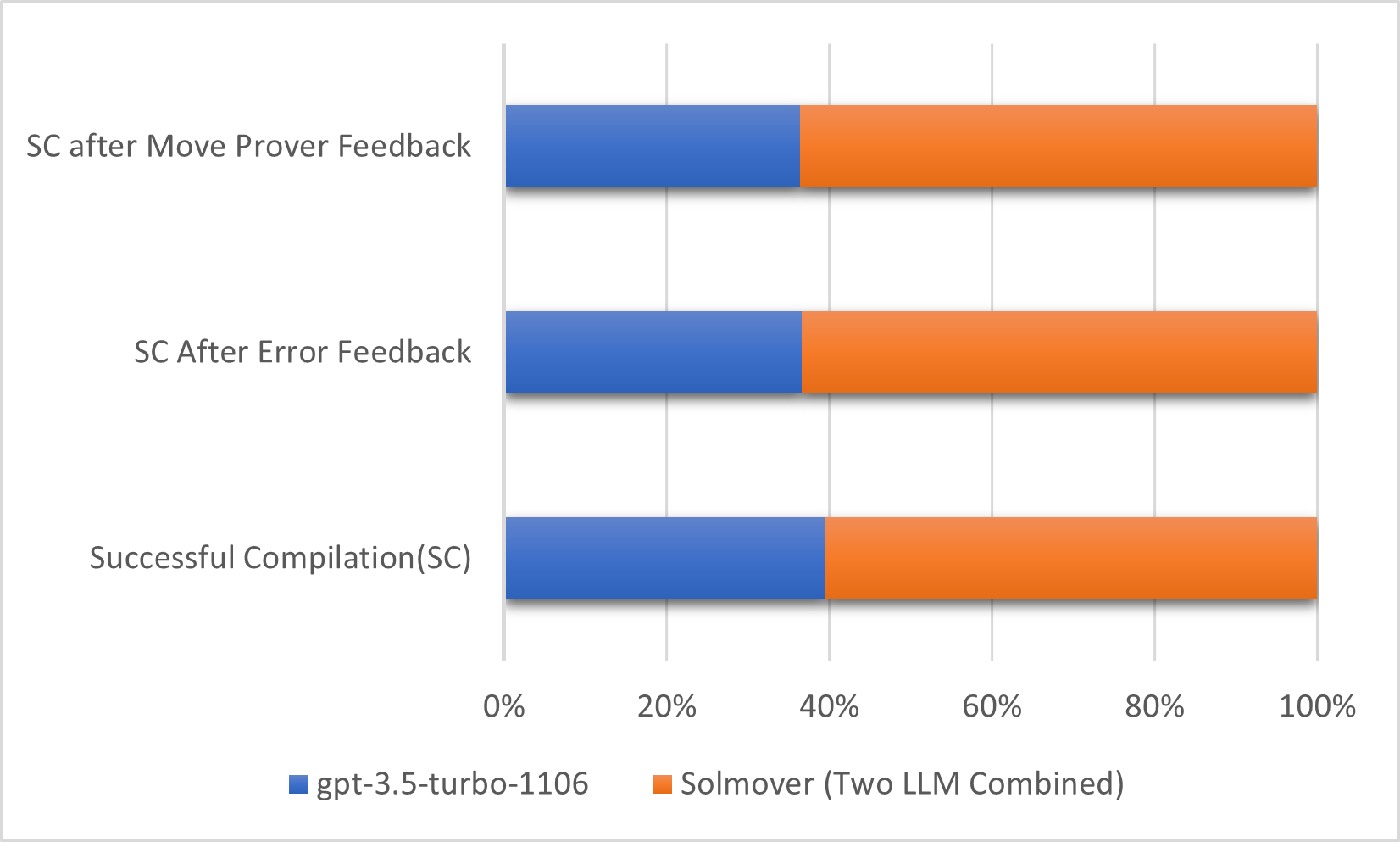}
    \caption{\label{translate3}Contract Correctness.}
    \label{fig:tran3}
\end{figure}

As we can see from Figure \ref{fig:tran3} that \texttt{Solmover} not only had a higher success rate in initial compilations but also showed greater improvement upon receiving feedback. Moreover, it continued to improve even after the Move Prover feedback, in contrast to the \texttt{gpt-3.5-turbo-1106}, which plateaued.

\begin{itemize}
    \item The \texttt{gpt-3.5-turbo-1106} had a baseline SC of 204, which increased to 229 after error feedback and remained unchanged after Move Prover feedback.
    \item The \texttt{Solmover} model, on the other hand, started with a higher SC of 313, which then improved to 397 after error feedback and slightly increased to 401 after Move Prover feedback.
\end{itemize}

The comparative assessment of \texttt{gpt-3.5-turbo-1106} and \texttt{Solmover (Two LLM Combined)} reveals that the latter outperforms the former in all stages of code compilation. Initially, \texttt{Solmover} starts with a higher number of successful compilations. It continues to improve significantly upon receiving error feedback and demonstrates a modest increase even after Move Prover feedback, suggesting a robust ability to learn and adapt. In contrast, \texttt{gpt-3.5-turbo-1106} exhibits improvement after error feedback but shows no further enhancement post Move Prover feedback, indicating a potential limit to its adaptability.

\section{Discussion \& Analysis}

In this section, we conduct analyses to understand the SolMover framework. We try to answer the research question (RQ) we started with and analyze the results reported.

\subsection{Concept Distillation}

A distinctive feature of our framework is the implementation of concept distillation via retrieval-augmented search methods. As illustrated in Figure \ref{fig:conceptdist}, our search methodology is adept at generating sub-tasks. A closer examination of the results presented in Figure 6, which addresses Contract Correctness, and Figure 5, focusing on Bug Mitigation, reveals the application of an iterative method to both the SolMover and the gpt-3.5-turbo-1106 model. Notably, the enhancement in performance and the reduction in the generation of non-compilable code are predominantly observed within our framework. This observation prompted a deeper investigation into our code translation strategy, particularly the generation of code using sub-tasks and the formulation of these sub-tasks through Concept Distillation.

\begin{tcolorbox}
\textbf{RQ1 \& RQ2} \textit{Are LLMs capable of learning coding concepts, and can these concepts facilitate the generation of detailed subtasks from a generalized prompt?}

Empirical evidence indicates a notable improvement in successful code translations when the model is prompted by sub-tasks derived from concepts. This leads to the conclusion that LLMs are capable of encoding associations that resemble concepts. However, we prefer to refer to these associations not as "concepts" in the traditional sense, but as "context associations."
\end{tcolorbox}

\subsection{Smart Contract Translation}

A central aim of this study was to assess the feasibility of translating smart contract code from one language (Solidity) to another (Move), a task complicated by the scarcity of Move code available on GitHub for model training purposes. We also explored various strategies to enhance the quality of the translated code by minimizing bugs introduced by the LLM. The findings, depicted in Figure 4, affirm the potential for successful translation of smart contracts from Solidity to Move, surpassing the performance of existing methodologies.

\begin{tcolorbox}
\textbf{RQ3} \textit{Is it possible for an LLM to generate code in a low-resource language it has not been explicitly trained on?}

Yes. Our findings suggest that, even in the absence of specific training in a given language, an LLM is capable of translating smart contracts from a well-documented language to a low-resource one, utilizing sub-tasks and minimal fine-tuning as facilitative measures.
\end{tcolorbox}

The incorporation of feedback from both the compiler and the prover into our iterative methodology contributes to a reduction in bugs and an enhancement in code correctness, as evidenced in Figures 6 and 5. Nevertheless, the improvement attributable to prover feedback appears to be minimal.

\begin{tcolorbox}
\textbf{RQ4} \textit{Does employing compiler feedback mitigate bugs?}

Yes. The empirical data indicates that iterative prompting with compiler feedback effectively reduces the incidence of uncompilable code. This positive outcome is observable across both models, with a more pronounced effect within our framework.
\end{tcolorbox}

\section{Related Work}

The related work can be categorized broadly as just Code Transpilers and LLM based solutions.
\textbf{Rule-based Transpilers.}
Rule-based transpilers serve as a conduit between different programming languages, akin to skilled interpreters. They operate on a set of predefined rules rather than conjecture, guaranteeing accurate and consistent code conversions. This method promotes collaboration and facilitates code reuse across languages, enabling straightforward transitions from Python to Java \cite{py2java} and Java to Python \cite{java2python}. Despite their precision and adaptability, rule-based transpilers are not without limitations. The establishment of conversion rules may necessitate a degree of programming acumen, and intricately featured language elements could require manual intervention for accurate translation.

\textbf{Statistical ML-based Transpilers.}
These transpilers harness machine learning to navigate the translation between programming languages. Diverging from rule-based strategies, statistical ML-based transpilers digest extensive corpora of parallel translations to discern patterns and correlations, aiding in the translation of novel code instances. Nguyen et al. \cite{nguyen2013lexical} and Karaivanov et al. \cite{karaivanov2014phrase} have contributed methodologies that blend phrase-based translation with an understanding of the grammatical structuring of the target language, augmented by custom rules. Aggarwal et al. \cite{aggarwal2015using} employed a similar strategy for transitioning from Python2 to Python3, leveraging sentence alignment techniques. Explorations into bidirectional transpilers, such as the one by Schultes \cite{schultes2021sequalsk} for Swift \cite{swift-lang} and Kotlin \cite{kotlin-lang}, and the CRustS transpiler by Ling et al. \cite{ling2022rust}, which emphasizes reducing unsafe expressions in the resultant code, represent significant advancements in this domain.

\textbf{Transformer and Other ML-based Code Translation Tools.}
Transformer models have revolutionized the field of code translation, drawing from their success in natural language processing. The encoder-decoder structure, as introduced by Vaswani et al. \cite{vaswani2017attention}, has profoundly influenced software engineering, inspiring the creation of specialized models for code translation. Variants such as CodeBERT \cite{feng2020codebert} and CodeGPT \cite{lu2021codexglue} have emerged, each with a focus either on code comprehension or the generation of coherent code in the target language. CodeT5 by Wang et al. \cite{wang2021codet5}, which integrates code semantics, and PLBART by Ahmad et al. \cite{ahmad2021unified}, a unified model designed for versatility across programming and natural languages, exemplify the ongoing innovation in this space. Efforts are also directed towards integrating developer insights and experimenting with novel architectures, including tree-to-tree models for program translation as pioneered by Chen et al. \cite{chen2018tree}.

\textbf{LLM-based Methods Using Compiler/unit testing.}
Unsupervised code translation techniques that capitalize on self-training and automated unit tests for verifying source-to-target code equivalence have been introduced by Roziere et al. \cite{roziere2022leveraging}. Although these methods utilize unit tests for creating synthetic parallel datasets for model enhancement, such tests are not factored into the training loss calculation. Wang et al. \cite{wang2022compilable}'s approach, which incorporates reinforcement learning with compiler feedback for code generation, offers a unique perspective distinct from supervised learning paradigms. Further research by Pan et al. \cite{haugeland2021migrating} and Orlanski et al. \cite{orlanski2023measuring} delves into the impact of LLMs on code translation, particularly in the context of low-resource languages, highlighting both the potential and challenges in this evolving field.

\section{Conclusion}

In this study, we introduce SolMover, an innovative framework comprising two Large Language Models (LLMs) designed to leverage textbook knowledge. Specifically, one LLM is utilized to distill and generate granular subtasks for the other, which then undertakes the task of generating code based on these subtasks. Our research focuses on the code translation task, with a particular emphasis on "Move" — a target language characterized by its extremely limited resources — for the creation of smart contracts.

The SolMover framework facilitates the translation of existing Solidity smart contracts into Move smart contracts, showcasing the feasibility of such translations even when the target language is significantly under-represented in training datasets. The contributions of our work are multifaceted:

\begin{itemize}
    \item We present a novel approach for encoding concepts within an LLM, illustrating how such encoded knowledge can assist in translating code into languages that the LLM was not explicitly trained on.
    \item Our exploration into an iterative compiler error feedback loop demonstrates its effectiveness in identifying and rectifying bugs within the translated code, thereby enhancing the overall quality and reliability of the output.
\end{itemize}
Through these contributions, SolMover advances the field of code translation, particularly in scenarios involving low-resource target languages, by demonstrating that it is possible to bridge the gap between different programming languages for smart contract development using LLMs. This approach not only expands the potential applications of LLMs in software engineering but also opens up new avenues for research in automated code translation and bug mitigation strategies.

\section*{Acknowledgement}
Part of this work was conducted with the generous grant support from Sui Foundation Academic Research Grant.

\bibliographystyle{unsrtnat}
\bibliography{references}  

\begin{thebibliography}{66}
\providecommand{\natexlab}[1]{#1}
\providecommand{\url}[1]{\texttt{#1}}
\expandafter\ifx\csname urlstyle\endcsname\relax
  \providecommand{\doi}[1]{doi: #1}\else
  \providecommand{\doi}{doi: \begingroup \urlstyle{rm}\Url}\fi

\bibitem[Liang et~al.(2022)Liang, Bommasani, Lee, Tsipras, Soylu, Yasunaga, Zhang, Narayanan, Wu, Kumar, et~al.]{liang2022holistic}
Percy Liang, Rishi Bommasani, Tony Lee, Dimitris Tsipras, Dilara Soylu, Michihiro Yasunaga, Yian Zhang, Deepak Narayanan, Yuhuai Wu, Ananya Kumar, et~al.
\newblock Holistic evaluation of language models.
\newblock \emph{ArXiv preprint}, abs/2211.09110, 2022.
\newblock URL \url{https://arxiv.org/abs/2211.09110}.

\bibitem[Bubeck et~al.(2023)Bubeck, Chandrasekaran, Eldan, Gehrke, Horvitz, Kamar, Lee, Lee, Li, Lundberg, et~al.]{bubeck2023sparks}
S{\'e}bastien Bubeck, Varun Chandrasekaran, Ronen Eldan, Johannes Gehrke, Eric Horvitz, Ece Kamar, Peter Lee, Yin~Tat Lee, Yuanzhi Li, Scott Lundberg, et~al.
\newblock Sparks of artificial general intelligence: Early experiments with gpt-4.
\newblock \emph{ArXiv preprint}, abs/2303.12712, 2023.
\newblock URL \url{https://arxiv.org/abs/2303.12712}.

\bibitem[Wu et~al.(2023)Wu, Wang, Wan, Jiao, and Lyu]{Wu2023ChatGPTOG}
Hao Wu, Wenxuan Wang, Yuxuan Wan, Wenxiang Jiao, and Michael~R. Lyu.
\newblock Chatgpt or grammarly? evaluating chatgpt on grammatical error correction benchmark.
\newblock \emph{ArXiv preprint}, abs/2303.13648, 2023.
\newblock URL \url{https://arxiv.org/abs/2303.13648}.

\bibitem[Moghaddam and Honey(2023)]{moghaddam2023boosting}
Shima~Rahimi Moghaddam and Christopher~J Honey.
\newblock Boosting theory-of-mind performance in large language models via prompting.
\newblock \emph{ArXiv preprint}, abs/2304.11490, 2023.
\newblock URL \url{https://arxiv.org/abs/2304.11490}.

\bibitem[Jiao et~al.(2023)Jiao, Wang, tse Huang, Wang, and Tu]{jiao2023ischatgpt}
Wenxiang Jiao, Wenxuan Wang, Jen tse Huang, Xing Wang, and Zhaopeng Tu.
\newblock Is chatgpt a good translator? a preliminary study.
\newblock In \emph{ArXiv}, 2023.

\bibitem[Agrawal et~al.(2023)Agrawal, Zhou, Lewis, Zettlemoyer, and Ghazvininejad]{agrawal-etal-2023-context}
Sweta Agrawal, Chunting Zhou, Mike Lewis, Luke Zettlemoyer, and Marjan Ghazvininejad.
\newblock In-context examples selection for machine translation.
\newblock In Anna Rogers, Jordan Boyd-Graber, and Naoaki Okazaki, editors, \emph{Findings of the Association for Computational Linguistics: ACL 2023}, pages 8857--8873, Toronto, Canada, 2023. Association for Computational Linguistics.
\newblock \doi{10.18653/v1/2023.findings-acl.564}.
\newblock URL \url{https://aclanthology.org/2023.findings-acl.564}.

\bibitem[Zhang et~al.(2023{\natexlab{a}})Zhang, Haddow, and Birch]{zhang2023prompting}
Biao Zhang, Barry Haddow, and Alexandra Birch.
\newblock Prompting large language model for machine translation: A case study.
\newblock \emph{ArXiv preprint}, abs/2301.07069, 2023{\natexlab{a}}.
\newblock URL \url{https://arxiv.org/abs/2301.07069}.

\bibitem[Vilar et~al.(2022)Vilar, Freitag, Cherry, Luo, Ratnakar, and Foster]{vilar2022prompting}
David Vilar, Markus Freitag, Colin Cherry, Jiaming Luo, Viresh Ratnakar, and George Foster.
\newblock Prompting palm for translation: Assessing strategies and performance.
\newblock \emph{ArXiv preprint}, abs/2211.09102, 2022.
\newblock URL \url{https://arxiv.org/abs/2211.09102}.

\bibitem[Lu et~al.(2023)Lu, Huang, Zhang, Yang, Lam, and Wei]{lu2023chain}
Hongyuan Lu, Haoyang Huang, Dongdong Zhang, Haoran Yang, Wai Lam, and Furu Wei.
\newblock Chain-of-dictionary prompting elicits translation in large language models.
\newblock \emph{ArXiv preprint}, abs/2305.06575, 2023.
\newblock URL \url{https://arxiv.org/abs/2305.06575}.

\bibitem[Bar-Hillel(1960)]{bar1960demonstration}
Yehoshua Bar-Hillel.
\newblock A demonstration of the nonfeasibility of fully automatic high quality translation.
\newblock \emph{Advances in computers}, 1:\penalty0 158--163, 1960.

\bibitem[Macklovitch(1995)]{macklovitch1995future}
Elliott Macklovitch.
\newblock The future of mt is now and bar-hillel was (almost entirely) right.
\newblock In \emph{Proceedings of the Fourth Bar-Ilan Symposium on the Foundations of Artificial Intelligence. url: http://rali. iro. umontreal. ca/Publications/urls/bisfai95. ps}, 1995.

\bibitem[Benson(2021)]{uniswap}
Jeff Benson.
\newblock Uniswap trading volume exploded by 450\% to \$7 billion. here’s why, 2021.
\newblock URL \url{https://decrypt.co/63280/uniswap-trading-volume-exploded-7-billion-heres-why}.

\bibitem[Rabimba et~al.(2022)Rabimba, Xu, Chen, Zhang, Gao, and Shi]{10.1145/3505253.3505259}
Karanjai Rabimba, Lei Xu, Lin Chen, Fengwei Zhang, Zhimin Gao, and Weidong Shi.
\newblock Lessons learned from blockchain applications of trusted execution environments and implications for future research.
\newblock In \emph{Workshop on Hardware and Architectural Support for Security and Privacy}, HASP '21, New York, NY, USA, 2022. Association for Computing Machinery.
\newblock ISBN 9781450396141.
\newblock \doi{10.1145/3505253.3505259}.
\newblock URL \url{https://doi.org/10.1145/3505253.3505259}.

\bibitem[Karanjai et~al.(2023{\natexlab{a}})Karanjai, Xu, Diallo, Chen, and Shi]{10174945}
Rabimba Karanjai, Lei Xu, Nour Diallo, Lin Chen, and Weidong Shi.
\newblock Defaas: Decentralized function-as-a-service for emerging dapps and web3.
\newblock In \emph{2023 IEEE International Conference on Blockchain and Cryptocurrency (ICBC)}, pages 1--3, 2023{\natexlab{a}}.
\newblock \doi{10.1109/ICBC56567.2023.10174945}.

\bibitem[Karanjai et~al.(2023{\natexlab{b}})Karanjai, Li, Xu, and Shi]{10316829}
Rabimba Karanjai, Edward Li, Lei Xu, and Weidong Shi.
\newblock Who is smarter? an empirical study of ai-based smart contract creation.
\newblock In \emph{2023 5th Conference on Blockchain Research \& Applications for Innovative Networks and Services (BRAINS)}, pages 1--8, 2023{\natexlab{b}}.
\newblock \doi{10.1109/BRAINS59668.2023.10316829}.

\bibitem[Zhong et~al.(2020)Zhong, Cheang, Qadeer, Grieskamp, Blackshear, Park, Zohar, Barrett, and Dill]{10.1007/978-3-030-53288-8_7}
Jingyi~Emma Zhong, Kevin Cheang, Shaz Qadeer, Wolfgang Grieskamp, Sam Blackshear, Junkil Park, Yoni Zohar, Clark Barrett, and David~L. Dill.
\newblock The move prover.
\newblock In Shuvendu~K. Lahiri and Chao Wang, editors, \emph{Computer Aided Verification}, pages 137--150, Cham, 2020. Springer International Publishing.
\newblock ISBN 978-3-030-53288-8.

\bibitem[Karanjai et~al.(2023{\natexlab{c}})Karanjai, Li, Xu, and Shi]{karanjai2023smarter}
Rabimba Karanjai, Edward Li, Lei Xu, and Weidong Shi.
\newblock Who is smarter? an empirical study of ai-based smart contract creation.
\newblock In \emph{2023 5th Conference on Blockchain Research \& Applications for Innovative Networks and Services (BRAINS)}, pages 1--8. IEEE, 2023{\natexlab{c}}.

\bibitem[Gile(2009)]{gile2009basic}
Daniel Gile.
\newblock \emph{Basic concepts and models for interpreter and translator training}.
\newblock Number~8 in Benjamins Translation Library. John Benjamins, Amsterdam, 2009.

\bibitem[Gunasekar et~al.(2023)Gunasekar, Zhang, Aneja, Mendes, Del~Giorno, Gopi, Javaheripi, Kauffmann, de~Rosa, Saarikivi, et~al.]{gunasekar2023textbooks}
Suriya Gunasekar, Yi~Zhang, Jyoti Aneja, Caio C{\'e}sar~Teodoro Mendes, Allie Del~Giorno, Sivakanth Gopi, Mojan Javaheripi, Piero Kauffmann, Gustavo de~Rosa, Olli Saarikivi, et~al.
\newblock Textbooks are all you need.
\newblock \emph{arXiv preprint arXiv:2306.11644}, 2023.

\bibitem[Zhang et~al.(2023{\natexlab{b}})Zhang, Muhamed, Anantharaman, Wang, Chen, Zhong, Cui, Xu, Zeng, Chilimbi, et~al.]{zhang2023reaugkd}
Jianyi Zhang, Aashiq Muhamed, Aditya Anantharaman, Guoyin Wang, Changyou Chen, Kai Zhong, Qingjun Cui, Yi~Xu, Belinda Zeng, Trishul Chilimbi, et~al.
\newblock Reaugkd: Retrieval-augmented knowledge distillation for pre-trained language models.
\newblock 2023{\natexlab{b}}.

\bibitem[Lu et~al.(2022)Lu, Duan, Han, Guo, Hwang, and Svyatkovskiy]{lu2022reacc}
Shuai Lu, Nan Duan, Hojae Han, Daya Guo, Seung-won Hwang, and Alexey Svyatkovskiy.
\newblock Reacc: A retrieval-augmented code completion framework.
\newblock \emph{arXiv preprint arXiv:2203.07722}, 2022.

\bibitem[Int(2023)]{IntroductionTheMoveBook-2023-08-12}
Introduction - the move book, 2023.
\newblock URL \url{https://move-language.github.io/move/}.

\bibitem[The(2022)]{TheMoveLanguageTheMoveBook-2022-11-25}
The move language - the move book, 2022.
\newblock URL \url{https://move-book.com/}.

\bibitem[mov(2023{\natexlab{a}})]{movepatterns}
Introduction - move patterns: Design patterns for resource based programming, 2023{\natexlab{a}}.
\newblock URL \url{https://www.move-patterns.com/}.

\bibitem[mov(2023{\natexlab{b}})]{movelanguagedocumentationtutorialatmainmovelanguagemoveGitHub-2023-12-19}
move · github, 2023{\natexlab{b}}.
\newblock URL \url{https://github.com/move-language/move/tree/main/ language/documentation/tutorial}.

\bibitem[Sui(2023)]{SuiBasicsSuiMovebyExample-2023-03-27}
Sui basics - sui move by example, 2023.
\newblock URL \url{https://examples.sui.io/basics/index.html}.

\bibitem[hfc(2023)]{hfcodegendatapreparedatasetpyatmainsayakpaulhfcodegenGitHub-2023-12-19}
hf-codegen, 2023.
\newblock URL \url{https://github.com/sayakpaul/ hf-codegen/blob/main/data/prepare_dataset.py}.

\bibitem[Karpukhin et~al.(2020)Karpukhin, O{\u{g}}uz, Min, Lewis, Wu, Edunov, Chen, and Yih]{karpukhin2020dense}
Vladimir Karpukhin, Barlas O{\u{g}}uz, Sewon Min, Patrick Lewis, Ledell Wu, Sergey Edunov, Danqi Chen, and Wen-tau Yih.
\newblock Dense passage retrieval for open-domain question answering.
\newblock \emph{arXiv preprint arXiv:2004.04906}, 2020.

\bibitem[Robertson et~al.(2009)Robertson, Zaragoza, et~al.]{robertson2009probabilistic}
Stephen Robertson, Hugo Zaragoza, et~al.
\newblock The probabilistic relevance framework: Bm25 and beyond.
\newblock \emph{Foundations and Trends{\textregistered} in Information Retrieval}, 3\penalty0 (4):\penalty0 333--389, 2009.

\bibitem[Oord et~al.(2018)Oord, Li, and Vinyals]{oord2018representation}
Aaron van~den Oord, Yazhe Li, and Oriol Vinyals.
\newblock Representation learning with contrastive predictive coding.
\newblock \emph{arXiv preprint arXiv:1807.03748}, 2018.

\bibitem[Taori et~al.(2023)Taori, Gulrajani, Zhang, Dubois, Li, Guestrin, Liang, and Hashimoto]{alpaca}
Rohan Taori, Ishaan Gulrajani, Tianyi Zhang, Yann Dubois, Xuechen Li, Carlos Guestrin, Percy Liang, and Tatsunori~B. Hashimoto.
\newblock Stanford alpaca: An instruction-following llama model.
\newblock \url{https://github.com/tatsu-lab/stanford_alpaca}, 2023.

\bibitem[Touvron et~al.(2023)Touvron, Lavril, Izacard, Martinet, Lachaux, Lacroix, Rozi{\`e}re, Goyal, Hambro, Azhar, et~al.]{touvron2023llama}
Hugo Touvron, Thibaut Lavril, Gautier Izacard, Xavier Martinet, Marie-Anne Lachaux, Timoth{\'e}e Lacroix, Baptiste Rozi{\`e}re, Naman Goyal, Eric Hambro, Faisal Azhar, et~al.
\newblock Llama: Open and efficient foundation language models.
\newblock \emph{arXiv preprint arXiv:2302.13971}, 2023.

\bibitem[ShareGPT(2023)]{sharegpt2023}
ShareGPT.
\newblock Sharegpt: Share your wildest chatgpt conversations with one click., 2023.
\newblock Available at: \url{https://sharegpt.com/}.

\bibitem[Urban(2004)]{urban2004mptp}
Josef Urban.
\newblock Mptp--motivation, implementation, first experiments.
\newblock \emph{Journal of Automated Reasoning}, 33:\penalty0 319--339, 2004.

\bibitem[fun(2023)]{funtoken}
fungible tokens, 2023.
\newblock URL \url{https://github.com/MystenLabs/sui/tree/main/sui\_programmability/examples/fungible\_tokens}.

\bibitem[bas(2023)]{basiccoin}
move/language/documentation/examples/experimental/basic-coin at main · move-language/move · github, 2023.
\newblock URL \url{https://github.com/move-language/move/tree/main/language/documentation/examples/experimental/basic-coin}.

\bibitem[die(2023)]{diem}
Diem, 2023.
\newblock URL \url{https://github.com/0LNetworkCommunity/libra-legacy-v6/blob/main/language/diem-framework/modules/Diem.move}.

\bibitem[tok(2023)]{token}
Token, 2023.
\newblock URL \url{https://github.com/starcoinorg/starcoin-framework/blob/main/sources/Token.move}.

\bibitem[gas(2023)]{gas}
Gas, 2023.
\newblock URL \url{https://github.com/0LNetworkCommunity/libra-legacy-v6/blob/main/language/diem-framework/modules/0L/GAS.move}.

\bibitem[nft(2023)]{nft}
Nft, 2023.
\newblock URL \url{https://github.com/MystenLabs/sui/tree/main/sui\_programmability/examples/nfts}.

\bibitem[mer(2023)]{merklenft}
starcoin-framework/sources/merklenft.move at main · starcoinorg/starcoin-framework · github, 2023.
\newblock URL \url{https://github.com/starcoinorg/starcoin-framework/blob/main/sources/MerkleNFT.move}.

\bibitem[def(2023)]{defi}
Defi, 2023.
\newblock URL \url{https://github.com/MystenLabs/sui/tree/main/sui\_programmability/examples/defi}.

\bibitem[coi(2023)]{coinswap}
move/language/documentation/examples/experimental/coin-swap at main · move-language/move · github, 2023.
\newblock URL \url{https://github.com/move-language/move/tree/main/language/documentation/examples/experimental/coin-swap}.

\bibitem[sta(2023)]{starswap}
Github - elements-studio/starswap-core: The swap project on starcoin such as uniswap a sushiswap, 2023.
\newblock URL \url{https://github.com/Elements-Studio/starswap-core}.

\bibitem[Ouyang et~al.(2022)Ouyang, Wu, Jiang, Almeida, Wainwright, Mishkin, Zhang, Agarwal, Slama, Ray, et~al.]{ouyang2022training}
Long Ouyang, Jeffrey Wu, Xu~Jiang, Diogo Almeida, Carroll Wainwright, Pamela Mishkin, Chong Zhang, Sandhini Agarwal, Katarina Slama, Alex Ray, et~al.
\newblock Training language models to follow instructions with human feedback.
\newblock \emph{Advances in Neural Information Processing Systems}, 35:\penalty0 27730--27744, 2022.

\bibitem[mix(2023)]{mixtral_moe}
Can you feel the moe? mixtral available with over 100 tokens per second through together platform!, 2023.
\newblock URL \url{https://www.together.ai/blog/mixtral}.

\bibitem[Hendy et~al.(2023)Hendy, Abdelrehim, Sharaf, Raunak, Gabr, Matsushita, Kim, Afify, and Awadalla]{hendy2023good}
Amr Hendy, Mohamed Abdelrehim, Amr Sharaf, Vikas Raunak, Mohamed Gabr, Hitokazu Matsushita, Young~Jin Kim, Mohamed Afify, and Hany~Hassan Awadalla.
\newblock How good are gpt models at machine translation? a comprehensive evaluation.
\newblock \emph{ArXiv preprint}, abs/2302.09210, 2023.
\newblock URL \url{https://arxiv.org/abs/2302.09210}.

\bibitem[py2()]{py2java}
{\texttt{py2java}: Python to Java Language Translator}.
\newblock \url{https://pypi.org/project/py2java/}.

\bibitem[Melhase et~al.()]{java2python}
Troy Melhase et~al.
\newblock {\texttt{java2python}: Simple but Effective Tool to Translate Java Source Code into Python}.
\newblock \url{https://github.com/natural/java2python}.

\bibitem[Nguyen et~al.(2013)Nguyen, Nguyen, and Nguyen]{nguyen2013lexical}
Anh~Tuan Nguyen, Tung~Thanh Nguyen, and Tien~N Nguyen.
\newblock {Lexical Statistical Machine Translation for Language Migration}.
\newblock In \emph{Proceedings of the 9th Joint Meeting on Foundations of Software Engineering}, pages 651--654, 2013.

\bibitem[Karaivanov et~al.(2014)Karaivanov, Raychev, and Vechev]{karaivanov2014phrase}
Svetoslav Karaivanov, Veselin Raychev, and Martin Vechev.
\newblock {Phrase-based Statistical Translation of Programming Languages}.
\newblock In \emph{{Proceedings of the 2014 ACM International Symposium on New Ideas, New Paradigms, and Reflections on Programming \& Software}}, pages 173--184, 2014.

\bibitem[Aggarwal et~al.(2015)Aggarwal, Salameh, and Hindle]{aggarwal2015using}
Karan Aggarwal, Mohammad Salameh, and Abram Hindle.
\newblock {Using Machine Translation for Converting Python 2 to Python 3 Code}.
\newblock Technical report, PeerJ PrePrints, 2015.

\bibitem[Schultes(2021)]{schultes2021sequalsk}
Dominik Schultes.
\newblock {SequalsK -- A Bidirectional Swift-Kotlin-Transpiler}.
\newblock In \emph{2021 IEEE/ACM 8th International Conference on Mobile Software Engineering and Systems (MobileSoft)}, pages 73--83. IEEE, 2021.

\bibitem[swi()]{swift-lang}
{Swift: The Powerful Programming Language that is Also Easy to Learn}.
\newblock \url{https://developer.apple.com/swift/}.

\bibitem[kot()]{kotlin-lang}
{Kotlin Programming Language: Concise. Cross‑platform. Fun}.
\newblock \url{https://kotlinlang.org/}.

\bibitem[Ling et~al.(2022)Ling, Yu, Wu, Wang, Cordy, and Hassan]{ling2022rust}
Michael Ling, Yijun Yu, Haitao Wu, Yuan Wang, James~R Cordy, and Ahmed~E Hassan.
\newblock {In Rust We Trust: A Transpiler from Unsafe C to Safer Rust}.
\newblock In \emph{{Proceedings of the ACM/IEEE 44th International Conference on Software Engineering: Companion Proceedings}}, pages 354--355, 2022.

\bibitem[Vaswani et~al.(2017)Vaswani, Shazeer, Parmar, Uszkoreit, Jones, Gomez, Kaiser, and Polosukhin]{vaswani2017attention}
Ashish Vaswani, Noam Shazeer, Niki Parmar, Jakob Uszkoreit, Llion Jones, Aidan~N Gomez, {\L}ukasz Kaiser, and Illia Polosukhin.
\newblock {Attention Is All You Need}.
\newblock \emph{{Advances in Neural Information Processing Systems (NeurIPS)}}, 30, 2017.

\bibitem[Feng et~al.(2020)Feng, Guo, Tang, Duan, Feng, Gong, Shou, Qin, Liu, Jiang, and Zhou]{feng2020codebert}
Zhangyin Feng, Daya Guo, Duyu Tang, Nan Duan, Xiaocheng Feng, Ming Gong, Linjun Shou, Bing Qin, Ting Liu, Daxin Jiang, and Ming Zhou.
\newblock {CodeBERT: A Pre-Trained Model for Programming and Natural Languages}.
\newblock In \emph{Findings of the Association for Computational Linguistics: EMNLP 2020}, pages 1536--1547, Online, November 2020. Association for Computational Linguistics.
\newblock \doi{10.18653/v1/2020.findings-emnlp.139}.

\bibitem[Lu et~al.(2021)Lu, Guo, Ren, Huang, Svyatkovskiy, Blanco, Clement, Drain, Jiang, Tang, et~al.]{lu2021codexglue}
Shuai Lu, Daya Guo, Shuo Ren, Junjie Huang, Alexey Svyatkovskiy, Ambrosio Blanco, Colin Clement, Dawn Drain, Daxin Jiang, Duyu Tang, et~al.
\newblock {CodeXGLUE: A Machine Learning Benchmark Dataset for Code Understanding and Generation}.
\newblock \emph{arXiv preprint arXiv:2102.04664}, 2021.

\bibitem[Wang et~al.(2021)Wang, Wang, Joty, and Hoi]{wang2021codet5}
Yue Wang, Weishi Wang, Shafiq Joty, and Steven~C.H. Hoi.
\newblock {CodeT5: Identifier-aware Unified Pre-trained Encoder-Decoder Models for Code Understanding and Generation}.
\newblock In \emph{{Proceedings of the 2021 Conference on Empirical Methods in Natural Language Processing}}, pages 8696--8708, Online and Punta Cana, Dominican Republic, 2021. Association for Computational Linguistics.
\newblock \doi{10.18653/v1/2021.emnlp-main.685}.
\newblock URL \url{https://aclanthology.org/2021.emnlp-main.685}.

\bibitem[Ahmad et~al.(2021)Ahmad, Chakraborty, Ray, and Chang]{ahmad2021unified}
Wasi Ahmad, Saikat Chakraborty, Baishakhi Ray, and Kai-Wei Chang.
\newblock {Unified Pre-training for Program Understanding and Generation}.
\newblock In \emph{Proceedings of the 2021 Conference of the North American Chapter of the Association for Computational Linguistics: Human Language Technologies}, pages 2655--2668, Online, June 2021. Association for Computational Linguistics.
\newblock \doi{10.18653/v1/2021.naacl-main.211}.
\newblock URL \url{https://aclanthology.org/2021.naacl-main.211}.

\bibitem[Chen et~al.(2018)Chen, Liu, and Song]{chen2018tree}
Xinyun Chen, Chang Liu, and Dawn Song.
\newblock {Tree-to-Tree Neural Networks for Program Translation}.
\newblock \emph{Advances in Neural Information Processing Systems (NeurIPS)}, 31, 2018.

\bibitem[Roziere et~al.(2022)Roziere, Zhang, Charton, Harman, Synnaeve, and Lample]{roziere2022leveraging}
Baptiste Roziere, Jie Zhang, Francois Charton, Mark Harman, Gabriel Synnaeve, and Guillaume Lample.
\newblock {TransCoder-ST: Leveraging Automated Unit Tests for Unsupervised Code Translation}.
\newblock In \emph{{International Conference on Learning Representations (ICLR)}}, 2022.
\newblock URL \url{https://openreview.net/forum?id=cmt-6KtR4c4}.

\bibitem[Wang et~al.(2022)Wang, Wang, Wan, Mi, Li, Zhou, Liu, Wu, Jiang, and Liu]{wang2022compilable}
Xin Wang, Yasheng Wang, Yao Wan, Fei Mi, Yitong Li, Pingyi Zhou, Jin Liu, Hao Wu, Xin Jiang, and Qun Liu.
\newblock {Compilable Neural Code Generation with Compiler Feedback}.
\newblock In \emph{{Findings of the Association for Computational Linguistics: ACL 2022}}, pages 9--19, 2022.

\bibitem[Haugeland et~al.(2021)Haugeland, Nguyen, Song, and Chauvel]{haugeland2021migrating}
Sindre~Gr{\o}nst{\o}l Haugeland, Phu~H Nguyen, Hui Song, and Franck Chauvel.
\newblock Migrating monoliths to microservices-based customizable multi-tenant cloud-native apps.
\newblock In \emph{2021 47th Euromicro Conference on Software Engineering and Advanced Applications (SEAA)}, pages 170--177. IEEE, 2021.

\bibitem[Orlanski et~al.(2023)Orlanski, Xiao, Garcia, Hui, Howland, Malmaud, Austin, Singh, and Catasta]{orlanski2023measuring}
Gabriel Orlanski, Kefan Xiao, Xavier Garcia, Jeffrey Hui, Joshua Howland, Jonathan Malmaud, Jacob Austin, Rishabh Singh, and Michele Catasta.
\newblock Measuring the impact of programming language distribution.
\newblock In \emph{International Conference on Machine Learning}, pages 26619--26645. PMLR, 2023.

\end{thebibliography}






\end{document}